\documentclass[epj]{svjour}
\usepackage{amssymb}
\usepackage{amsmath}
\usepackage{graphicx}
\usepackage{dcolumn}
\usepackage{bm}
\usepackage{graphics}
\begin{document}
\title{Delay-induced cluster patterns in coupled Cayley tree networks}
\author{Aradhana Singh and Sarika Jalan}
\institute{Complex Systems Lab,
Indian institute of technology Indore, IET-DAVV campus Khandwa Road, Indore-452017 India }
\date{Received: date / Revised version: date}

\abstract{We study effects of delay in diffusively coupled logistic maps on
the Cayley tree networks. We find that smaller coupling values exhibit sensitiveness
to value of delay, and lead to different cluster patterns of self-organized and driven
types. Whereas larger coupling strengths exhibit robustness against change in delay values,
and lead to stable driven clusters comprising nodes from last generation of the Calaye tree.
Furthermore, introduction of delay exhibits suppression as well as enhancement of
synchronization depending upon coupling strength values. To the end we discuss the importance of
results to understand conflicts and cooperations observed
in family business.
}

\PACS{
{05.45.Xt}{ } \and {05.45.Pq}{} 
}
\maketitle

\section{Introduction}
Many real-world networks display local 
co-ordination among nodes leading to cluster synchronization 
\cite{Nature2010,Science2010,SJ_prl2003,Kurths_prl2006}. 
Formation of clusters, which are based on the 
dynamical properties of the coupled system, 
typically depends on the
underlying network structure.
The interplay between the structure and the dynamics of complex networks has been a 
focus of intense research 
interest in last decades \cite{functional_net}. 
Furthermore, delay naturally arises in extended systems due to the finite
speed of information transmission \cite{book_delay}. 
For example, in neural networks, propagation delays of electrical 
signals connecting different neurons and
local neurovascular couplings lead to time delays 
\cite{book_neural1,book_neural3}.
A delay may give rise to many new phenomena in dynamical systems such
as oscillation death, enhancement or suppression of synchronization, chimera state, etc
\cite{osc_death_delay,delay_supress_syn,delay_enhance_syn,delay_coup_osc,chimera}.
The existence of delay can completely change the behavior of a system as 
observed for undelayed case \cite{book_delay}.
What follows that time delay might be deliberately implemented in order 
to achieve desired functions such as 
secure communication \cite{secure_comm} 
and to control neural disturbances, e.g., suppression of
undesired synchrony of firing neurons in Parkinson's disease or epilepsy 
\cite{neural_disease_delay1,neural_disease_delay2,neural_disease_delay3}.

Our recent work demonstrated that delay plays a crucial role in formation of
synchronized clusters and mechanism behind the synchronization.
We  presented results for cluster formation in 
delayed coupled maps on 1-d lattice,
small-world, scale-free, random and complete bipartite networks \cite{Singh2012}.
In this paper we investigate delay-induced cluster patterns in
diffusively coupled logistic maps on Cayley tree networks.

The Cayley tree is an infinite dimensional regular
graph with an idealized hierarchical structure \cite{rev_network}.
Its rich hierarchal structure turns out to be an ideal model network to investigate
driven patterns in details. Furthermore, regularity of Cayley trees makes analytical understanding
or origin of driven patterns easier to understand using Lyapunov function analysis.
 
Cayley trees provide a simple model to do exact analysis for stability of synchronized 
states \cite{CML_tree}, to study localization criteria in impurity atom \cite{solving_prob},
to derive expression for magnetization and zero field 
susceptibility \cite{cayley_Ising}, etc.. Biologically oriented work on Cayley tree networks
include modeling of immune networks with 
antibody dynamics \cite{cayley_immune}. In a recent paper, Cayley trees have been used to 
investigate Bose-Einstein condensation \cite{cayley_bose-condensation}.

\section{Model}
We use well known delayed coupled maps model \cite{Singh2012}:
\begin{equation}
x_i(t+1) = (1-\varepsilon) f(x_i(t)) + \frac{\varepsilon}{k_i} \sum_{j=1}^N A_{ij} g(x_j(t - \tau))
\label{cml}
\end{equation}
Here
$ k_{i}$ = $\sum_{j=1}^{N}A_{ij}$ is degree, and 
$x_i(t)$ is the dynamical variable of the
$i-th$ node ($1 \le i \le N$) at time $t$, $A$ is the 
adjacency matrix with elements $A_{ij}$ taking values
$1$ and $0$ depending upon whether there is a 
connection between $i$ and $j$ or not. 
The delay $\tau$ is the time it takes for the information 
from a unit to reach its neighbors and be processed.
The function $f(x)$ defines the local nonlinear map 
and the function $g(x)$ defines the nature
of coupling between the nodes. We present the 
results for the local dynamics given by the logistic map
$f(x) = \mu x (1 - x)$ and for diffusive coupling $g(x)
=f(x)$. We take $\mu = 4$, for which logistic map
exhibits chaotic behavior. 

\section{Phase synchronization and synchronized clusters}
Synchronization of coupled dynamical systems 
is defined by the appearance of some relation between the functional 
of different dynamical variables. The exact synchronization corresponds to the
situation where the dynamical variables for different
 nodes have identical values. The phase synchronization 
corresponds to the situation where the dynamical variables
 for different nodes have some definite relation 
between their phases \cite{phase_syn}. 
 We consider
phase synchronization as defined in \cite{SJ_prl2003,phase_syn_prl1998}.

{\it Phase synchronized clusters:}
Let $n_i$ and $n_j$ denote the number of times
the variables $x_i(t)$ and $x_j(t)$, $t=1,2, 
\hdots T$ for the nodes $i$ and $j$, show local minima
during the time interval $T$. Let $n_{ij}$ 
denote the number of times these local minima match with 
each other. Phase distance between two 
nodes $i$ and $j$ is given as $d_{ij} = 1 - 2n_{ij}/(n_i + n_j)$.
Clearly, $d_{ij} = 0$ when all minima of 
variables $x_i$ and $x_j$ match with each other $d_{ij} = 1$ when
none of the minima match. Nodes $i$ and 
$j$ are phase synchronized if $d_{ij}=0$. 
A cluster of nodes is
phase synchronized if all pairs of nodes of the cluster are phase synchronized.

\begin{figure}
\includegraphics[height=1.5in,width=3.5in]{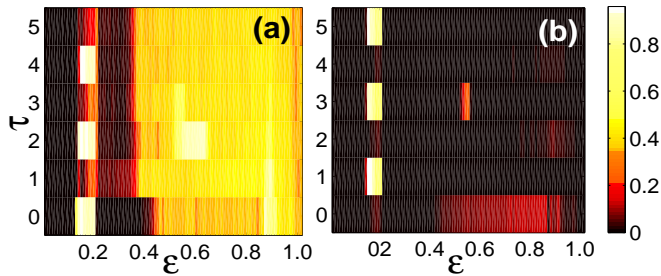}
\caption{(Color online) Phase diagram demonstrating different 
values of (a) $f_{inter}$ and (b) $f_{intra}$ in two parameter space
of $\varepsilon$ and $\tau$ 
((a) and (b) ) with $N=127$, $<k>=2$. Local
dynamics is governed by logistic map $f(x)=4x(1-x)$ and coupling function
$g(x)=f(x)$.
The figure is obtained by averaging over 20 random initial conditions. 
The color-scale encoding represents values of $f_{inter}$ and $f_{intra}$.
The regions, which are black in both graphs
(a) and (b), correspond to states of no cluster formation.
The regions, where both subfigures have gray shades (yellow), correspond to
states where clusters with both inter- and 
intra-couplings are formed. The regions in (a), which are lighter as compared to
the corresponding $\varepsilon$ and $\tau$ values in (b),
refer to dominant D phase synchronized clusters states,
and the reverse refer to dominant SO phase
 synchronized clusters state. White (light yellow) regions
in (a) and (b) refer to ideal D and ideal SO clusters state respectively.
The regions, which are dark gray in (a) and black in (b) or vice-versa, correspond to states where
very few nodes are forming the cluster.}
\label{Fig_Tree_Phase}
\end{figure}

\section{Mechanisms of cluster formation}
Depending upon the asymptotic dynamical behavior the nodes 
of the network can be divided into the following three types \cite{pre2005a}. 
\\
 (a) {\it Cluster node} synchronizes with
other nodes and forms a synchronized cluster. Once this node
enters a synchronized cluster it remains in that cluster afterwards.
\\
 (b) {\it Isolated node} does not synchronized
 with any other node and remains isolated all the time.
\\
 (c) {\it Floating node} keeps on switching 
intermittently between an independent evolution and a
synchronized evolution attached to a cluster.

The study of  
relation between the synchronized clusters and the coupling
between the nodes represented by the adjacency matrix 
exhibits following two different phenomena of cluster formation.
\\
(1) {\it Self-organized clusters}: The nodes of a cluster can be
synchronized because of intra-cluster couplings. 
We refer to this as the self-organized (SO) synchronization and the corresponding 
synchronized clusters as SO clusters. Ideal SO synchronization refers to a
state when clusters do not have
any connection outside the cluster, except one. 
Dominant SO synchronization corresponds 
to the state when most of the connections lie inside the cluster.
\\
͑(2) {\it Driven clusters}: The nodes of a cluster can be 
synchronized because of inter-cluster couplings. 
We refer this as driven (D) synchronization and
the corresponding cluster as D cluster. 
The ideal D synchronization refers to the state
when clusters do not have any connections within them, and all connections are
outside.
Dominant D synchronization corresponds to 
the state when most of the connections lie outside 
the cluster and very few inside.
 
To get a clear picture of self-organized and driven 
behavior we consider two quantities
$f_{intra}$ and $f_{inter}$ as measures for 
intra-cluster and inter-cluster couplings as follows:
\begin{equation}
f_{intra} = \frac{N_{intra}}{N_c}, \,\, \, \, \, \, \, f_{inter} = \frac{N_{inter}}{N_c}
\end{equation}
where $N_{intra}$ and $N_{inter}$ are the numbers 
of intra- and inter- cluster couplings, respectively. In 
$N_{inter}$, coupling between two isolated nodes are not included.

We define one more important state which forms basic backbone of
the present investigation.\\
{\it Cluster patterns}: A cluster pattern refers to a particular phase synchronized state, 
which contains information of all the pairs of phase synchronized nodes
distributed in various clusters.
A cluster pattern can be static or dynamical.
Static pattern has all nodes fixed, except few floating nodes,
in a cluster with respect to change in time, delay value or initial condition. 
Dynamical pattern exhibits changes with time evolution, or 
with initial condition or 
with change in delay value. A change in the pattern refers to the state when members of a cluster get
changed.
Furthermore, patterns can be of D or SO type, which
respectively refers to a particular D or SO phase synchronized state.

\begin{figure}
\centerline{\includegraphics[width=0.65\columnwidth]{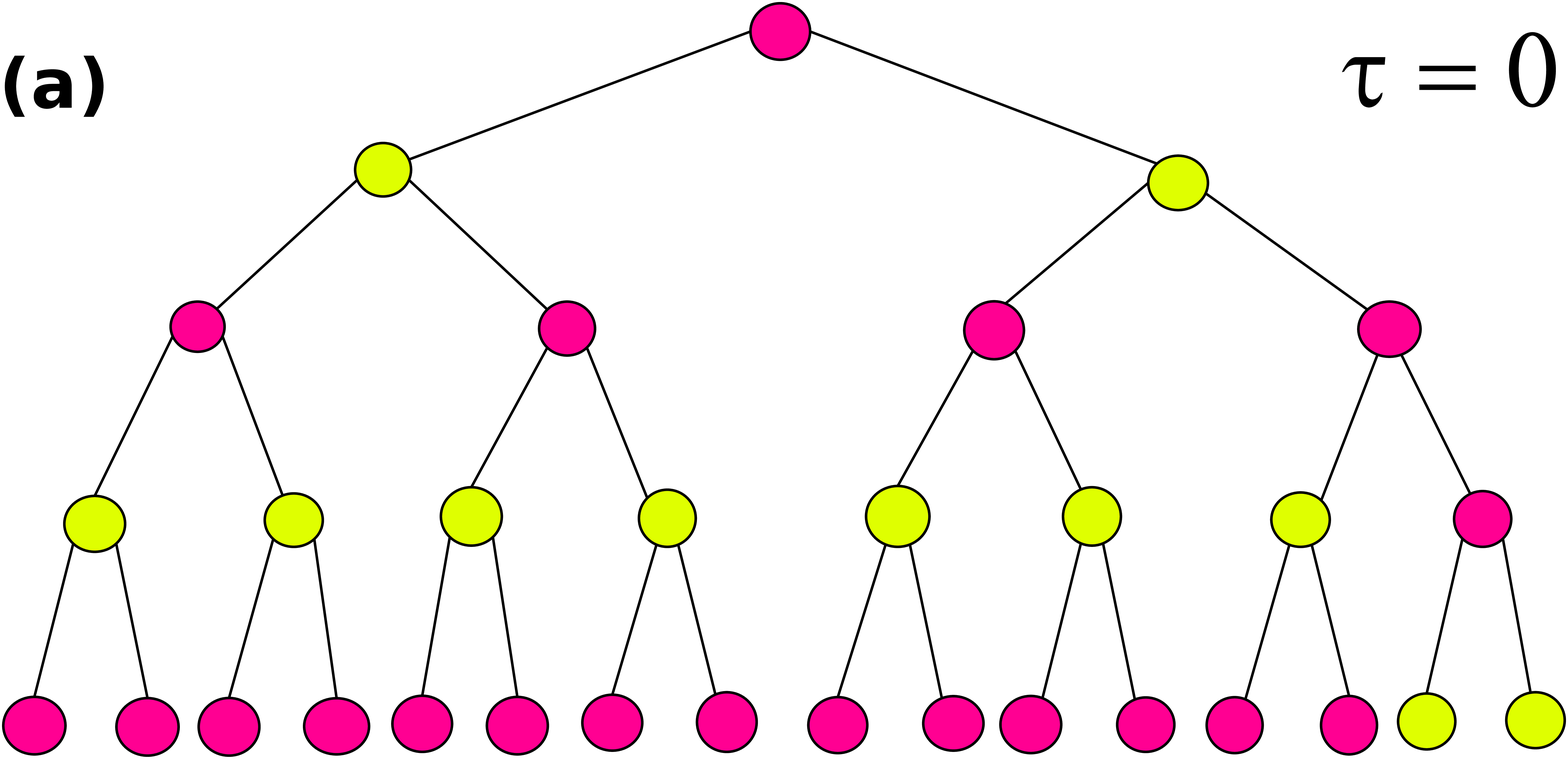}}
\centerline{\includegraphics[width=0.65\columnwidth]{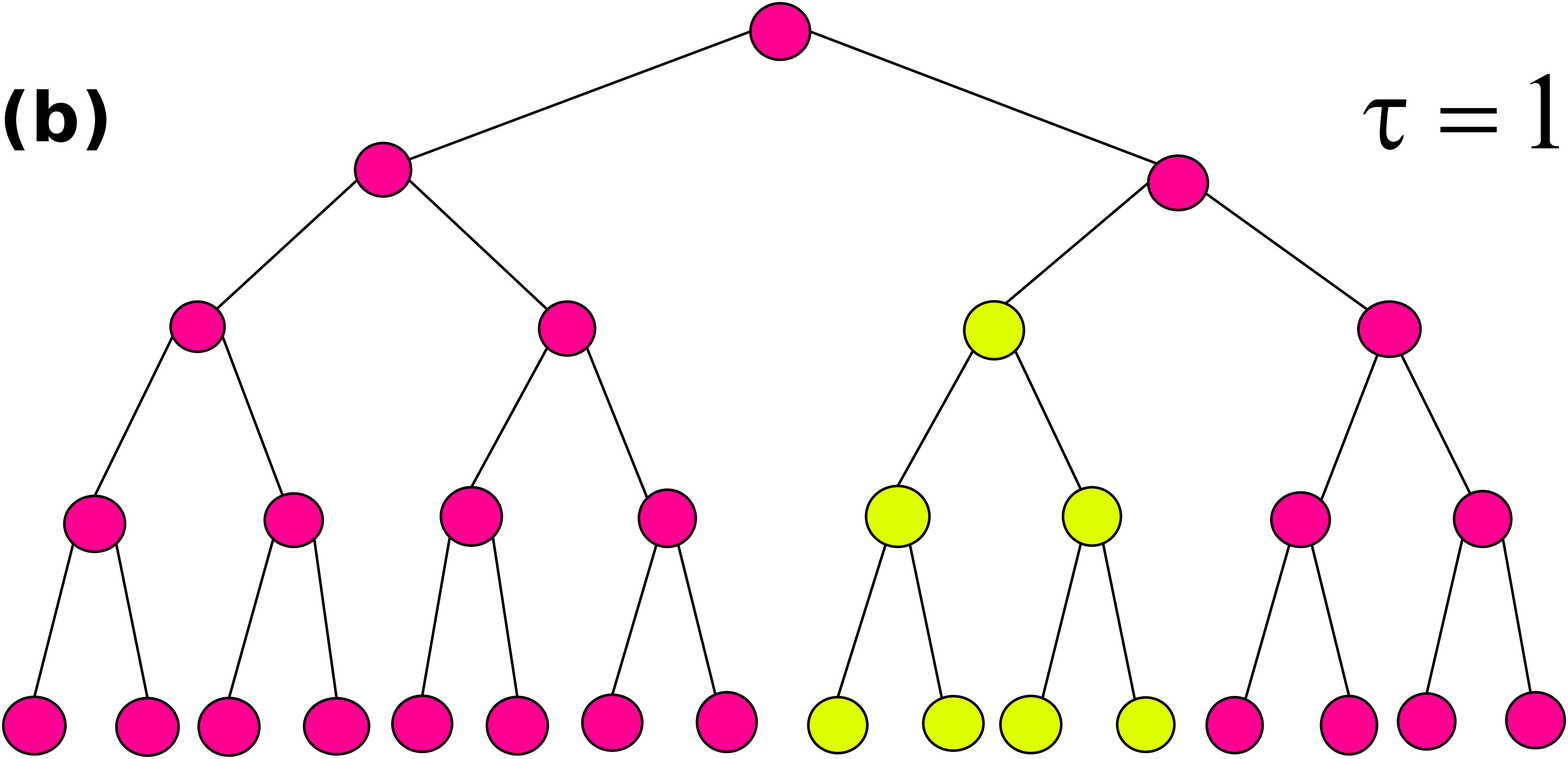}}
\centerline{\includegraphics[width=0.65\columnwidth]{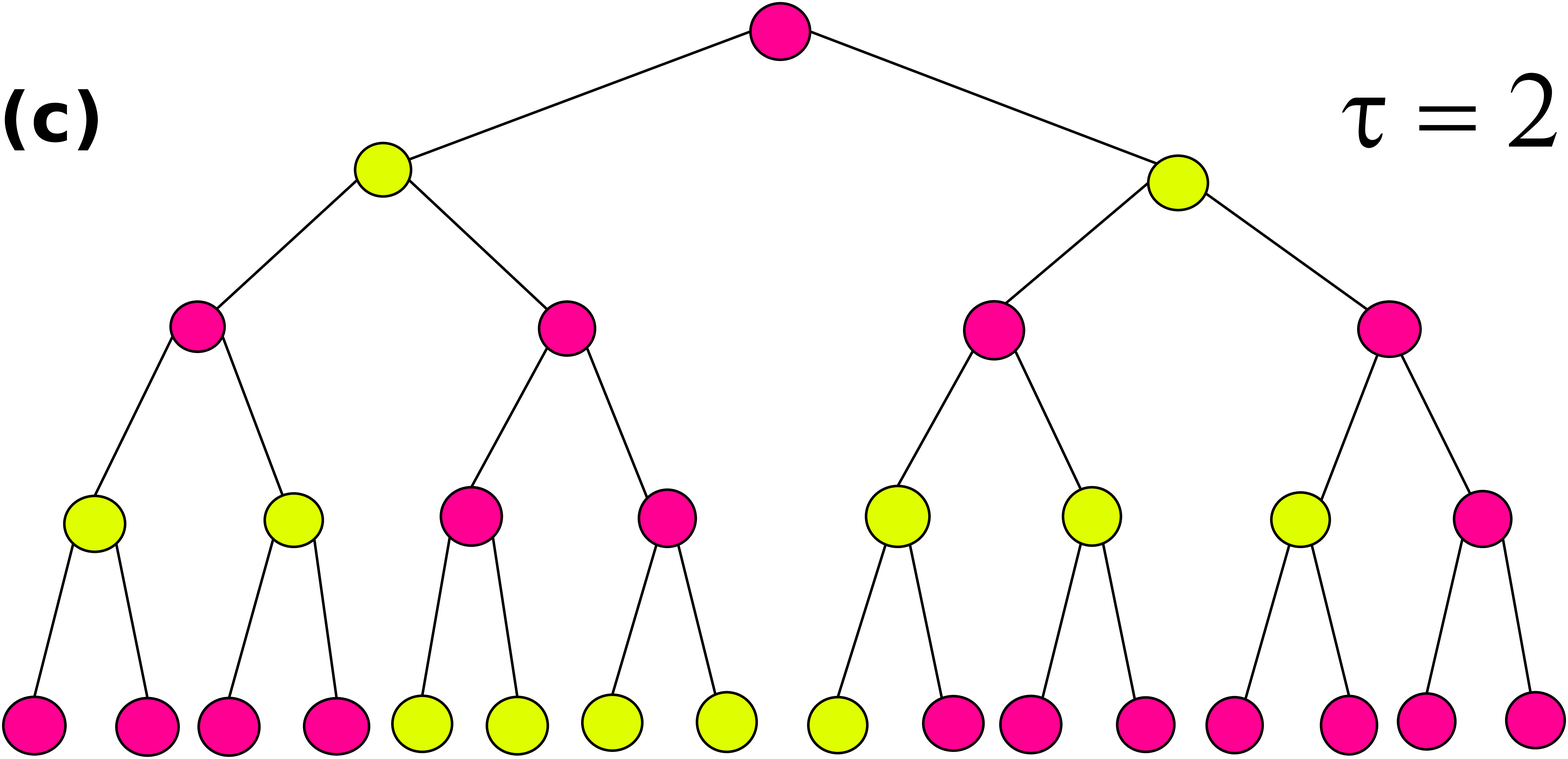}}
\centerline{\includegraphics[width=0.65\columnwidth]{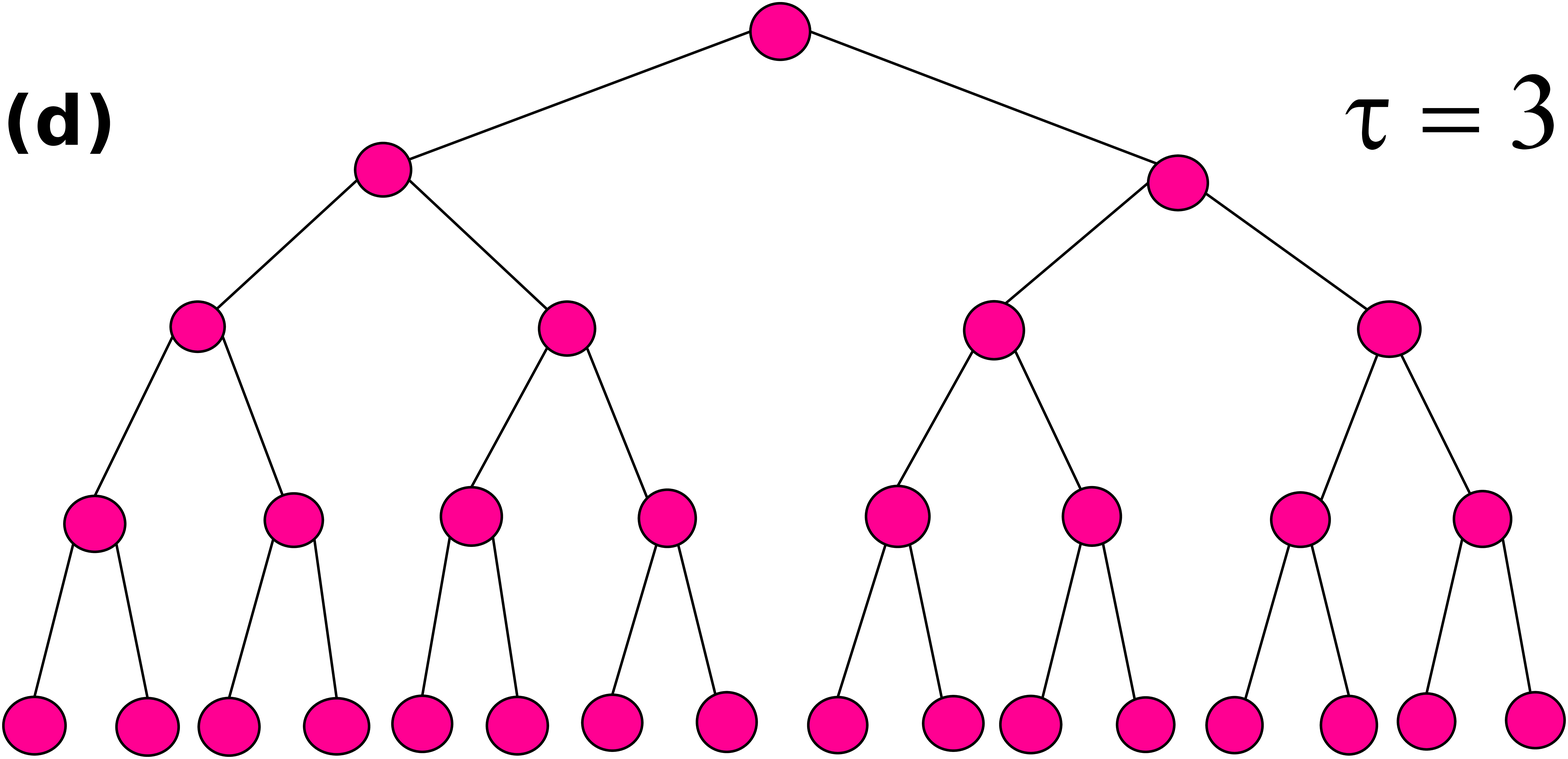}}
\centerline{\includegraphics[width=0.65\columnwidth]{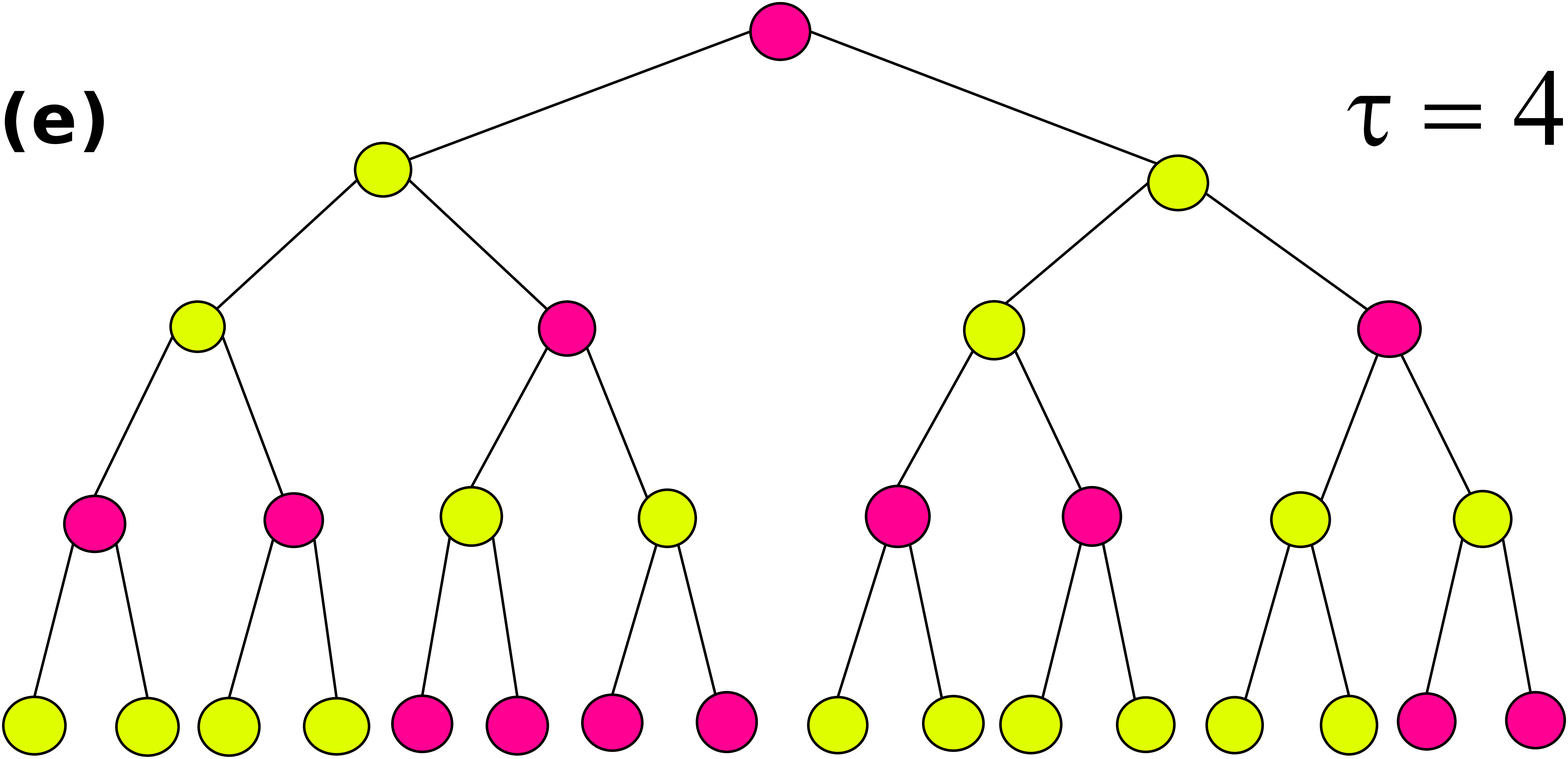}}
\centerline{\includegraphics[width=0.65\columnwidth]{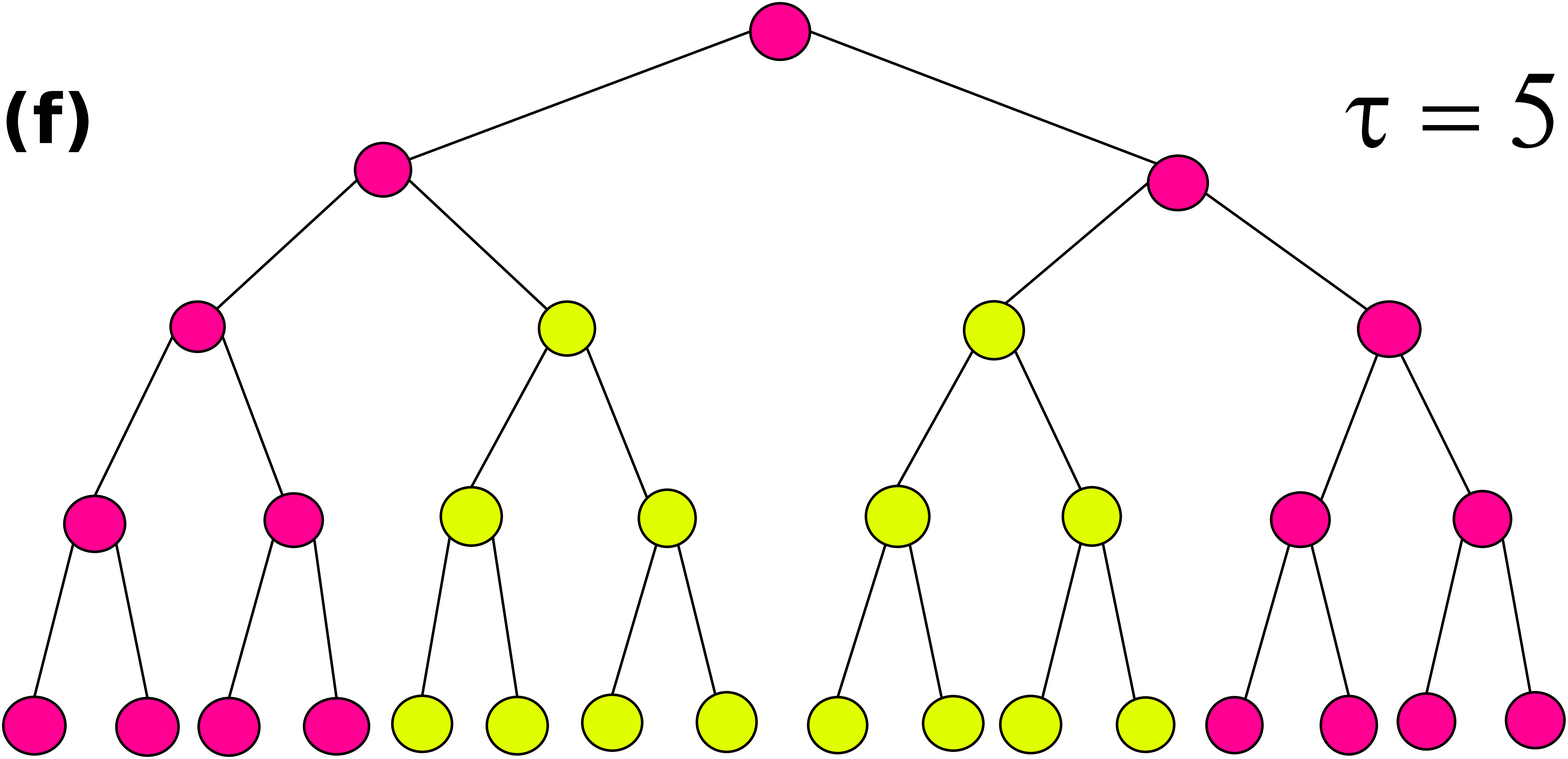}}
\caption{(Color online) Schematic network diagrams illustrating different
cluster patterns observed for different delay values at lower coupling strength region.
The examples are for  $N=31, <k>=2$ and $\varepsilon=0.16$.
The closed circles of same shade (color) imply that the 
corresponding nodes are phase synchronized (i.e. $C_{ij}=1$),
and the open circles imply that the corresponding nodes are  not phase
synchronized.The D chaotic clusters for $\tau=0,2$ and $4$.
The SO  periodic clusters for $\tau=1$, $3$ and $5$. }
\label{Fig_Tree_change_mechanism}
\end{figure}

\section{Numerical Results}
We evolve Equation~(\ref{cml}) starting from random initial 
conditions, and study the dynamical
behavior of nodes after an initial transient. 
We study phase synchronized clusters for $100$ time 
steps after the initial transient, and calculate values 
of $f_{inter}$ and $f_{intra}$ as described earlier.

Phase diagrams Figs.(\ref{Fig_Tree_Phase}a) and (\ref{Fig_Tree_Phase}b)
are plotted for network size $N=127$ and average degree $<k>=2$.
Undelayed Cayley tree exhibits dominant D clusters
in the range $0.12 \lesssim \varepsilon \lesssim 0.19$ with a
periodic dynamical evolution.
With a further increase in coupling strength, there is no phase synchronization till 
$\varepsilon = 0.4$, after which dominant D clusters are obtained as elucidated by  
light gray (yellow) regions in Fig.~(\ref{Fig_Tree_Phase}a) and dark gray (red) 
regions in Fig.~(\ref{Fig_Tree_Phase}b). At very high coupling strengths, persistence of
light gray (yellow) regions in (\ref{Fig_Tree_Phase}a) and appearance of
black regions in  (\ref{Fig_Tree_Phase}b) indicate ideal D clusters.

On introduction of a delay $\tau=1$ in the evolution equation Eq.~\ref{cml}, 
after very small coupling values for which there is no phase synchronization
(black color for the subfigures ~(\ref{Fig_Tree_Phase}a) 
and (\ref{Fig_Tree_Phase}b)), SO phase synchronized clusters are formed for 
$0.12 {\lesssim} \varepsilon \lesssim 0.19$ as elucidated by white 
regions in (\ref{Fig_Tree_Phase}b). 
As coupling strength increases, in the range 
$0.36 \lesssim \varepsilon \lesssim 0.42$,
where undelayed system exhibits no or very less cluster formation,
delayed evolution manifests dominant D clusters as depicted by gray (yellow) regions in 
Fig.~(\ref{Fig_Tree_Phase}a). With a further increase in coupling strength, appearance 
of gray (yellow) regions in Fig.~(\ref{Fig_Tree_Phase}a) and black regions in 
Fig.~(\ref{Fig_Tree_Phase}b) indicate ideal D clusters.

For $\tau=2$, appearance of white (light yellow) window in Fig.~(\ref{Fig_Tree_Phase}a) 
and corresponding window in (\ref{Fig_Tree_Phase}b) with dark gray (red) to black shades
indicate formation of dominant and ideal D 
clusters respectively in lower coupling values. This description is same as observed for 
$\tau=0$.
Larger coupling strengths lead to ideal D clusters as
elucidated by black and light gray (yellow) regions 
in Figs.~(\ref{Fig_Tree_Phase}b) and (\ref{Fig_Tree_Phase}a) respectively.

For a further increase in delay,  at lower coupling strength
odd delay values exhibit similar behavior as observed for $\tau=1$, while even delay
values manifest similar behavior as observed for $\tau=0$ and  $\tau=2$.
At higher coupling values, coupled dynamics for all delays demonstrate 
either ideal or dominant D clusters.
\begin{figure}
\centerline{\includegraphics[width=3.2in, height=1.0in]{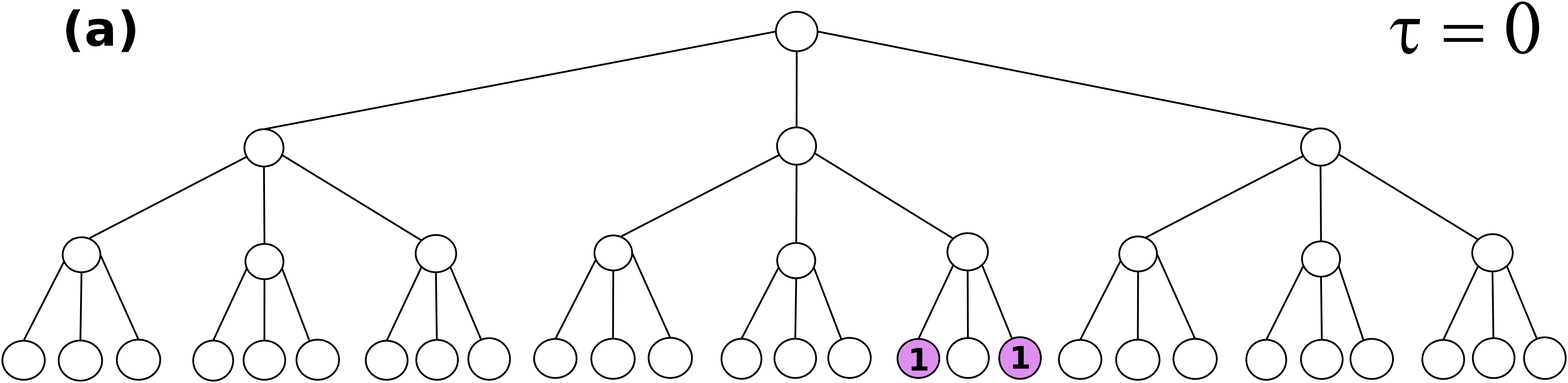}}
\centerline{\includegraphics[width=3.2in, height=1.0in]{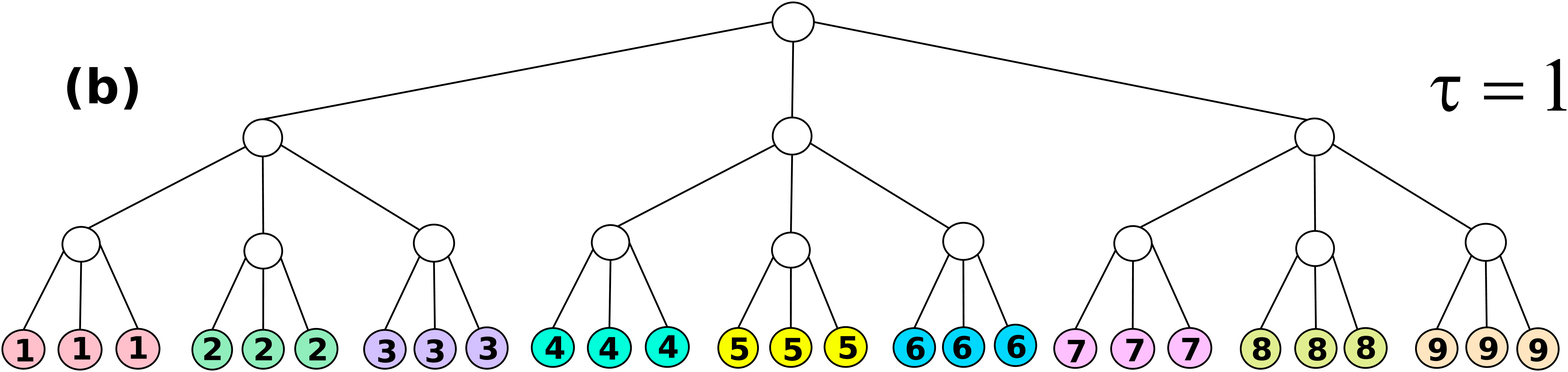}}
\centerline{\includegraphics[width=3.2in, height=1.0in]{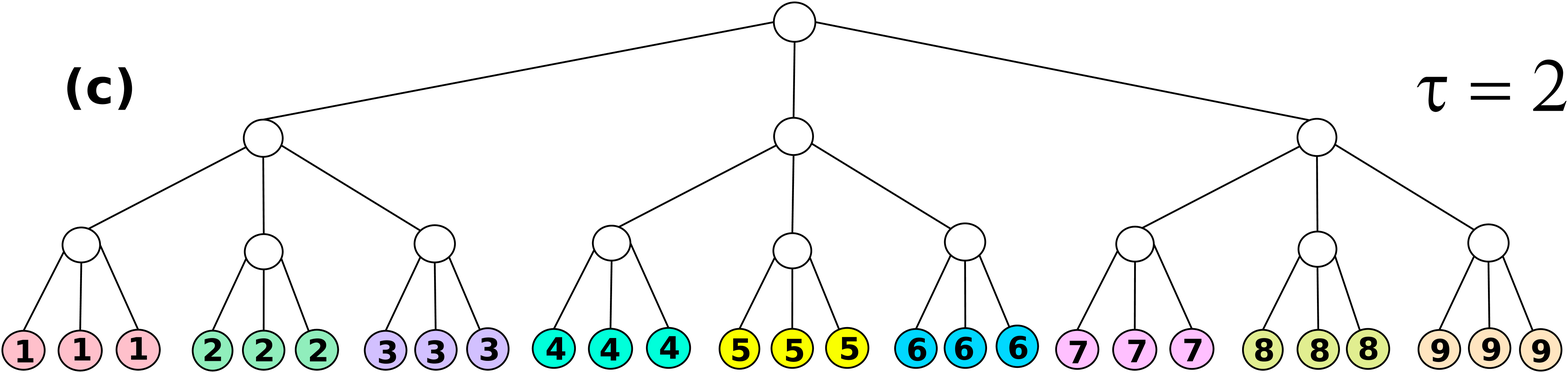}}
\caption{(Color online) Schematic diagrams illustrating delay-induced driven 
patterns for larger coupling values. 
 The examples are for  $N=40, <k>=3$ and $\varepsilon=0.37$.
 The closed circles of same number (same color) imply that the 
corresponding nodes are phase synchronized (i.e. $A_{ij}=1$),
and the open circles imply that the corresponding nodes are  not phase
synchronized.The D chaotic clusters for $\tau=1, 2$ }
\label{Fig_Tree_pattern1}
\end{figure}
\begin{figure}
\centerline{\includegraphics[width=3.2in, height=1.0in]{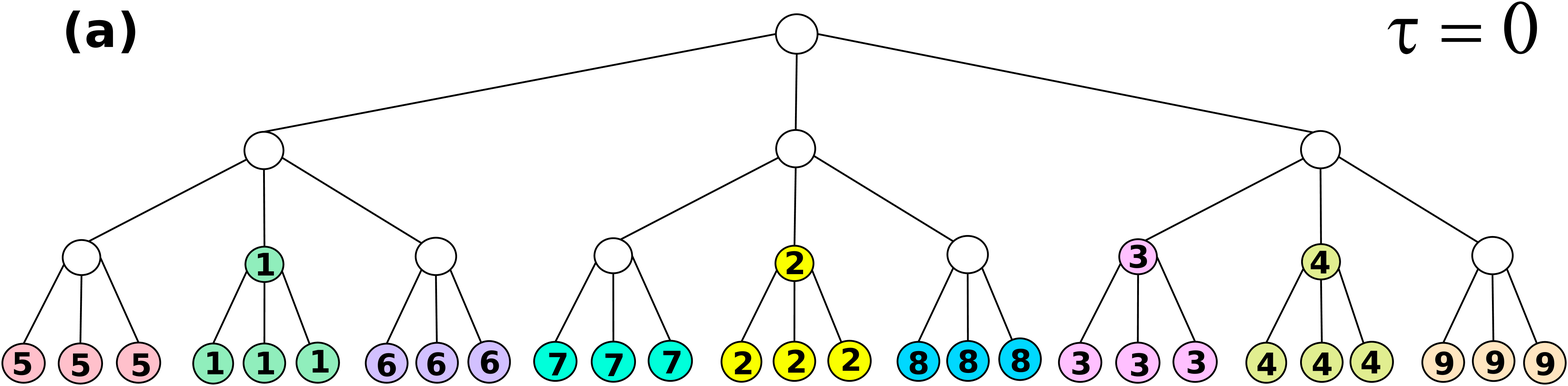}}
\centerline{\includegraphics[width=3.2in, height=1.0in]{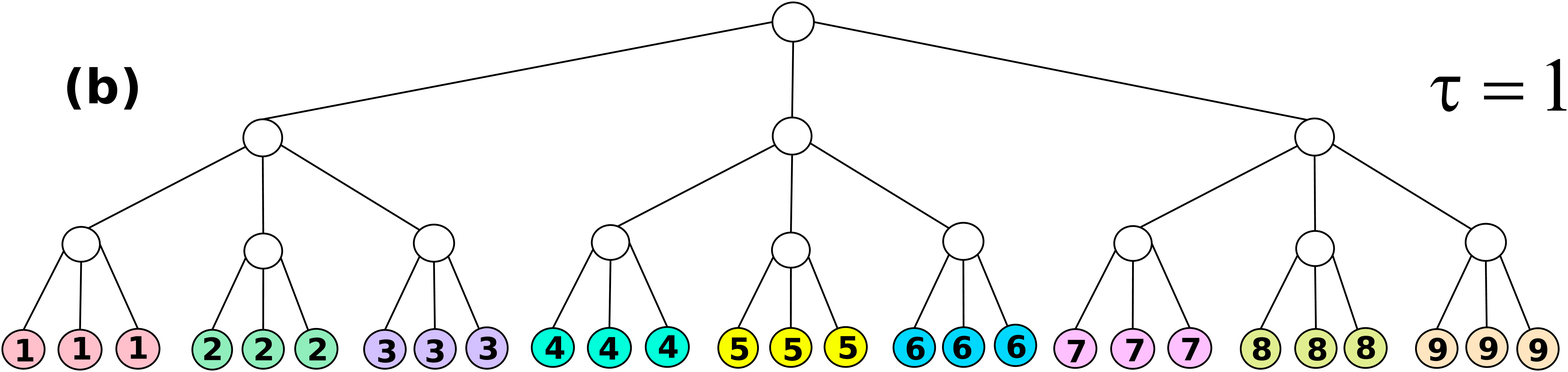}}
\caption{(Color online) Schematic diagrams illustrating
effects on delay on phase synchronized patterns.
The examples are for  $N=30, <k>=3$ and $\varepsilon=0.7$.
Closed circles having same number (same color) imply that  
corresponding nodes are phase synchronized in same cluster,
and open circles imply that corresponding nodes are not phase
synchronized.}
\label{Fig_Tree_pattern2}
\end{figure}

\subsection{Delay-induced change in mechanism of cluster formation:}
Above discussions indicate that at the lower coupling values, 
change in delay values are related with the change in mechanism behind the cluster formation.
Odd delay values lead to ideal or dominant SO clusters, whereas  
even delay values are associated with ideal or dominant D clusters.
Figs.~(\ref{Fig_Tree_change_mechanism}a), (\ref{Fig_Tree_change_mechanism}c) and 
(\ref{Fig_Tree_change_mechanism}e) illustrate that
for $\tau=0$, $\tau=2$ and $\tau=4$, nodes in alternate generations are synchronized
with each other,  except few cases where nodes in two consecutive generations (parents
and children) too exhibit synchronization. 
Figs.~(\ref{Fig_Tree_change_mechanism}b), (\ref{Fig_Tree_change_mechanism}d) and 
(\ref{Fig_Tree_change_mechanism}f) demonstrate that for $\tau=1$, $\tau=3$ and $\tau=5$,  
either one single cluster is formed spanning all nodes,
or several clusters are formed with clusters consisting of 
nodes in consecutive generations.

The cluster patterns observed here are dynamical with respect to intimal condition as well
as with delay value,
but for a particular value of delay the phenomenon 
behind the synchronization in cluster-pattern is static and 
same parity of delay leads to same phenomenon of cluster synchronization.
The dynamical evolution in this range of 
coupling strength is periodic for all delay values.

\subsection{Delay-induced driven patterns}
As described in earlier sections, for coupling 
range $0.35 \lesssim \varepsilon {\lesssim} 0.42$, 
undelayed coupled maps do not exhibit cluster formation, whereas delayed evolution 
leads to dominant D clusters. In order to explain different these dynamical cluster patterns 
clearly we make schematic diagram of dynamical clusters in 
Fig.~(\ref{Fig_Tree_pattern1}). For undelayed evolution there is no co-ordination between
any pair of nodes, and hence there is no cluster pattern as depicted by all empty circles
in Fig.~(\ref{Fig_Tree_pattern1}a).  
Introduction of delay induces co-ordination 
between nodes in the same sub-family of the last generation, as depicted by different clusters
in Fig.(~\ref{Fig_Tree_pattern1}b). 
Further change in delay value does not have any measure impact on synchronized clusters state.
These cluster patterns are stable
with respect to time evolution, 
initial condition as well as change in delay value. 

For a further increase in $\varepsilon$,
delay destroys the co-ordination between the nodes which are connected, giving rise to ideal 
D patterns with again only last generation nodes being synchronized in several clusters
(Fig.(~\ref{Fig_Tree_pattern2}b)). 
These patterns are too stable with respect to change in initial 
condition or change in delay value.
Furthermore, delayed as well undelayed
dynamical evolution in this range are associated with the chaotic evolution.

\section{Lyapunov function analysis} 
As demonstrated above, while undelayed evolution in middle coupling strength exhibits
some synchrony between children in last generation and their parents, and hence
giving rise to SO clusters, delayed evolution yields only D cluster indicating
loss of synchrony between parent and children.
Introduction of delay destroys the synchronization between parents and children while
keeping the co-ordination between children unaffected.
In order to understand this behavior let us write down Lyapunov function for a pair of 
synchronized nodes as,
\begin{eqnarray}
& & V_{ij}(t+1) = [ (1-\epsilon)( f(x_i(t)) - f(x_j(t))) + \nonumber\\
&  & \frac{ \varepsilon}{\sum_{k=1}^N A_{ij}} \sum_{k=1}^N A_{ik}{g}(x_k(t - \tau))  - \nonumber\\ 
& & \frac{  \varepsilon}{\sum_{k=1}^N A_{ij}} \sum_{k=1}^{N}A_{ik} {g}(x_k(t - \tau)) ]^{2}
\nonumber
\end{eqnarray}
Lyapunov function takes following simple form for two nodes originated from the same parent node,
  \begin{eqnarray}
& & V_{ij}(t+1) = [ (1-\epsilon)( f(x_i(t)) - f(x_j(t))) ]^{2}
\nonumber
\end{eqnarray}
Above equation does not have any delay term. All this leads to the conclusion that   
if a pair of nodes originated from same parent are synchronized, 
the introduction of delay would not affect the synchrony between them.
Introduction of delay only affects the connected nodes such as parent and children, 
as it removes the common term in their evolution equations,
and hence may be a reason behind the destruction of 
co-ordination between them \cite{Singh2012}.

\section{Conclusion}
We have studied delay-induced patterns in coupled maps on Cayley tree networks.
We demonstrate that different delay values manifest different cluster 
patterns at lower coupling values, where change in delay not only leads to a completely new pattern 
but also associated with different mechanisms behind the cluster formation.
In middle coupling range, delayed evolution always exhibits D clusters. Though we always
find either ideal or dominant D clusters in this range, the role of delay on evolution 
and co-ordinations among nodes are very different for different coupling values. 
For
some coupling values where undelayed evolution does not exhibit any synchronization, 
a introduction of delay enhances synchronization between children in last generation
yielding D clusters, whereas for some coupling values
delay destroys existing synchrony between parents and children while keeping
children coordinated, again giving rise to D clusters.
These delay induced D clusters are stable with respect to 
the change in delay value and consists only last generation nodes.
Lyapunov function analysis provides some hints about formation of stable D clusters in this region.

The model which we have considered here demonstrates that lower coupling
strength in general favors synchronization in various generations in the family, as indicated
by larger cluster size and by large number of nodes (almost all) participating in clusters. 
These clusters are sensitive with respect to external conditions such a time delay. Where as higher 
coupling strength leads to very drastic behavior such as formation of stable D clusters comprising
only last generations. 
The origin of these stable D clusters for last
generation nodes can be very well understood for Calaye three, where coupling 
environment of the last generation nodes belonging to same sub-family remains same,
and hence gives rise to a stable driven cluster \cite{Singh2012}, whereas nodes
originated from the same parent in any previous generation can not have same coupling 
environment unless their all children are synchronized with each other. 

Coupling strengths can be interpreted as closeness or bonding 
among family members, for example lower coupling strengths can correspond to a situation
where members live in nuclear families and do not share much details apart that
they belong to a same big family. Where as larger coupling strengths can be treated as a situation
where all members of a family live together as a joint family \cite{joint_family_India}. 
Our results may be used to understand 
conflicts in brothers running a successful family business  
\cite{family_business}, which on 
very simple terms can be attributed to the conflicts between their children (as shown in
Figs.(3) and (4)), where as lower coupling strength keeps
a warmer relation leading to cooperation in family (as seen for Fig.~(2)). Lower coupling 
strength here can be considered
as separating the business of siblings and cousins, which have been proven
to increase cooperation between them \cite{family_business_conflict}.

To conclude, we demonstrate that delay in spatially extended systems
may lead to a 
completely different relation between the 
functional clusters and topology than exhibited by undelayed evolution, and hence provide
an additional step towards ongoing research attributed to  understand relation between these two.
Since delay has already been emphasized to be important for
many real world networks \cite{book_delay}, the results presented in the paper is important to understand
various different behaviors exhibited by these systems. Furthermore, observation of different
cluster patterns as a function of delay may shed some light in understanding conflicts
or cooperation in family business \cite{family_business_conflict_book}.

\section*{Acknowledgment}SJ thanks DST for financial support.

\end{document}